\documentclass[12pt,a4paper,notitlepage]{article}

\usepackage{amssymb,amsmath}
\usepackage[dvips]{graphicx}

\newcommand{\bsl}{\boldsymbol}
\newcommand{\rcm}{\boldsymbol{R}_{\text{cm}}}

\begin{document}

\title{\bf Ground-state baryons in nonperturbative quark dynamics}

\author{I.M.Narodetskii and M.A.Trusov \\
{\itshape ITEP, Moscow, Russia}}

\maketitle \vspace{1cm}

\begin{abstract}\noindent
We review the results obtained  in an Effective Hamiltonian (EH)
approach for the three-quark systems. The EH is derived starting
from the Feynman--Schwinger representation for the gauge-invariant
Green function of the three quarks propagating in the
nonperturbative QCD vacuum and assuming the minimal area law for
the asymptotic of the Wilson loop. It furnishes the QCD consistent
framework within which to study baryons. The EH has the form of
the nonrelativistic three-quark Hamiltonian with the perturbative
Coulomb-like and nonperturbative string interactions and the
specific mass term. After outlining the approach, methods of
calculations of the baryon eigenenergies and some simple
applications are explained in details. With only two parameters:
the string tension $\sigma=0.15\text{~GeV}^2$ and the strong
coupling constant $\alpha_s=0.39$ a unified quantitative
description of the ground state light and heavy baryons is
achieved. The prediction of masses of the doubly heavy baryons not
discovered yet are also given. In particular, a mass of
$3660\text{~MeV}$ for the lightest $\Xi_{cc}$ baryon is found by
employing the hyperspherical formalism to the three quark
confining potential with the string junction.
\end{abstract}

\newpage

\section{Introduction}
The heavy flavor physics started with the discovery of the charm
quark in 1974. It was followed by the discovery of the beauty
quark  in 1977 and of the top  quark  in 1995. Soon after the
discovery of the charm quark several charmed baryons and mesons
have been identified. The discovery of single charmed or beauty
baryons has been followed by the discovery of the  $B_c$ meson
\cite{ABE} and recently by the extensive search of doubly charmed
baryons \cite{Mattson02}.

Doubly heavy baryons are baryons that contain two heavy quarks,
either $cc$, $bc$, or $bb$. Their existence is a natural
consequence of the quark model of hadrons, and it would be
surprising if they did not exist. In particular, data from the
BaBar and Belle collaborations at the SLAC and KEK B-factories
would be good places to look for doubly charmed baryons. Recently
the SELEX, the charm hadroproduction experiment at Fermilab,
reported a narrow state at $3519\pm 1\text{~MeV}$ decaying in
$\Lambda_c^+K^-\pi^+$, consistent with the weak decay of the
doubly charmed baryon $\Xi_{cc}^+$ \cite{Mattson02}. The candidate
is $6.3\sigma$ signal.

The SELEX result was recently critically discussed in \cite{KL02}.
Whether or not the state that SELEX reports turns out to be the
first observation of doubly charmed baryons, studying their
properties is important for a full understanding of the strong
interaction between quarks.

Estimations for the masses and spectra of the baryons containing
two or more heavy quarks have been considered by many authors
\cite{BaBar}. The purpose of this paper is to present a consistent
treatment of the results of the calculation\footnote{The preview
of this calculation has been done in \cite{NT01}.} of the masses
and wave functions of the doubly heavy baryons obtained in a
simple approximation within the nonperturbative QCD. As a
by-product, we also report the masses and wave functions for light
and heavy baryons.

The paper is organized as follows. In Section 2 we briefly review
the Effective Hamiltonian method. In Section 3 we discuss the
hyperspherical approach which is a very effective numerical tool
to solve this Hamiltonian. In Section 4 our predictions for the
ground-state spectra of light and heavy baryons are reported and a
detailed comparison of spectra of doubly heavy baryons with the
results of other approaches is given. Section 5 contains our
conclusions.

\section{The Effective Hamiltonian in QCD}

Starting from the QCD Lagrangian and assuming the minimal area law
for the asymptotic of the Wilson loop, the Hamiltonian of the 3q
system in the rest frame has been derived. The methodology of the
approach has been reviewed recently \cite{Si02} and so will be
sketched here only briefly. The Y-shaped baryon wave function has
the form:
\begin{equation}
B_Y(x_1,x_2,x_3,X) = e_{\alpha\beta\gamma}q^\alpha(x_1,X)
q^\beta(x_2,X)q^\gamma(x_3,X), \label{baryon} \end{equation} where
$\alpha$, $\beta$, $\gamma$ are the color indices, $q(x_i,X)$ is
the extended operator of the $i^{\text{th}}$ quark at a point
$x_i$,  and $X=(0,\bsl{X})$ is the equilibrium junction position.
The wave function (\ref{baryon}) describes the only
gauge-invariant configuration possible for baryons. The starting
point of the approach is the Feynman--Schwinger representation for
the gauge-invariant Green function of the three quarks propagating
in the nonperturbative QCD vacuum
\begin{equation}
\label{Greenfunction} G(x,y)=\prod\limits_{i=1}^3
\int\limits^\infty_0 ds_i \int Dz_i \exp (-K_i) \langle
\mathcal{W}\rangle_B,
\end{equation} where $x=\{x_1,x_2,x_3\}$, $y=\{y_1,y_2,y_3\}$, $z_i=z_i(s_i)$ are
the quark trajectories with
$z_i(0)=x_i$, $z_i(T)=y_i$, while $s_i$ is the Fock--Schwinger
proper time of the $i^{\text{th}}$ quark. Angular brackets mean
averaging over background field. The quantities $K_i$ are the
kinetic energies of quarks, and all the dependence on the vacuum
background field is contained in the generalized Wilson loop
$\mathcal{W}$
\begin{equation}
\mathcal{W}=\frac{1}{3!}\varepsilon_{ijk}\varepsilon_{lmn}U_1^{il}U_2^{jm}U_3^{kn},
\end{equation}
with
\begin{equation}U_k=P\exp(ig\int\limits_{C_i}A_{\mu}(x)dx^{\mu}),~~~k=1,2,3.\end{equation}
Here $P$ denotes the path-ordered product along the path $C_i$ in
Fig. 1 where the contours run over the classical trajectories of
static quarks. In this figure three quark lines start at junction
$X$ at time zero, run in the time direction from $0$ to $T$ with
the spatial position of quarks fixed and join again in the
junction $Y$ at time $T$. There are three planes that are bounded,
respectively, by one quark line, two lines connecting junction and
quark at $t=0$ and $t=T$, and the connection line of two
junctions. Under the minimal area law assumption, the Wilson loop
configuration takes the form
\begin{equation} \langle {\cal W}\rangle_B \propto\exp
(-\sigma(S_1+S_2+S_3)), \label{area} \end{equation} where $S_i$
are the minimal areas inside the contours formed by quarks and the
string junction  trajectories and $\sigma$ is the QCD string
tension.

In Eq. (\ref{Greenfunction}) the role of the time parameter along
the trajectory of each quark is played by the Fock--Schwinger
proper time $s_i$. The proper and real times for each quark are
related via a new quantity that eventually plays the role of the
dynamical quark mass. The final result is the derivation of the
Effective Hamiltonian (EH), see Eq. (\ref{EH}) below.

In contrast to the standard approach of the constituent quark
model, the dynamical masses $m_i$ are no longer free parameters.
They are expressed in terms of the running masses $m^{(0)}_i(Q^2)$
defined at the appropriate hadronic scale of $Q^2$ from the
condition of the minimum of the baryon mass as a function of
$m_i$.

Technically, this has been done using the einbein (auxiliary
fields) approach, which is proven to be rather accurate in various
calculations for relativistic systems. Einbeins are treated as
$c$-number variational parameters: the eigenvalues of the EH are
minimized with respect to einbeins to obtain the physical
spectrum. Such procedure, first suggested in \cite{FS91},
\cite{mesons1}, provides the reasonable accuracy for the meson
ground states \cite{mesons2}.

The method was already applied to study baryon Regge trajectories
\cite{FS91} and very recently for computation of magnetic moments
of light baryons \cite{KS00}. The essential point adopted in
\cite{NT01} and continued in this paper is that it is very
reasonable that the same method should also give a unified
description for both light and heavy baryons  including baryons
with two heavy quarks. As in \cite{KS00} we take as the universal
QCD parameter the string tension $\sigma$. We also include the
perturbative Coulomb interaction with the frozen strong coupling
constant $\alpha_s$.

From experimental point of view, a detailed discussion of the
excited $QQ'q$ states is probably premature. Therefore we consider
the ground-state baryons without radial and orbital excitations,
in which case tensor and spin-orbit forces do not contribute
perturbatively. Then only the spin-spin interaction survives in
the perturbative approximation. In what follows we disregard the
hyper-fine splitting among the baryon masses, then the EH has the
following form:
\begin{equation}
\label{EH} H=\sum\limits_{i=1}^3\left(\frac{{m_i^{(0)}}^2}{2m_i}+
\frac{m_i}{2}\right)+H_0+V.
\end{equation}
Here $H_0$ is the nonrelativistic kinetic energy operator and $V$
is the sum of the perturbative one-gluon-exchange potential $V_C$:
\[
V_C=-\frac{2}{3}\alpha_s\cdot\sum\limits_{i<j}\frac{1}{r_{ij}},
\]
where $r_{ij}$ are the distances between quarks, and the string
potential $V_{\text{string}}$. The baryon mass is given by formula
\begin{equation}
M_B=\min\limits_{m_i} \langle H\rangle +C,
\end{equation}
where $C$ is the quark self-energy correction calculated in
\cite{Si01}:
\begin{equation} \label{self_energy}
C=-\frac{2\sigma}{\pi}\sum\limits_i\frac{\eta_i}{m_i}
\end{equation}
with $\eta=1$ for $q$-quark\footnote{Here and throughout the paper
$q$ denotes a light quark $u$ or $d$}, $\eta=0.88$ for $s$-quark,
$\eta=0.234$ for $c$-quark, and $\eta=0.052$ for $b$-quark.

The string potential calculated in \cite{FS91} as the static
energy of the three heavy quarks was shown to be consistent with
that given by a minimum length configuration of the strings
meeting in a Y--shaped configuration at a junction $\bsl{X}$:
\begin{equation}
V_{\text{string}}(\bsl{r}_1,\bsl{r}_2, \bsl{r}_3)=\sigma
l_{\text{min}},\end{equation}  where $l_{\text{min}}$ is the sum
of the three distances between quarks and the string junction
point $\bsl{X}$. The Y--shaped  configuration was suggested long
ago \cite{ADM}, and since then was used repeatedly in many
dynamical calculations \cite{CKP83}.

\section{Solving the three-quark equation}

\subsection{Jacobi coordinates}

The baryon wave function depends on the three-body Jacobi
coordinates \begin{equation}\label{rho}
\bsl{\rho}_{ij}=\sqrt{\frac{\mu_{ij}}{\mu}}(\bsl{r}_i-\bsl{r}_j),
\end{equation}\begin{equation}\label{lambda}\bsl{\lambda}_{ij}=\sqrt{\frac{\mu_{ij,k}}{\mu}}
\left(\frac{m_i\bsl{r}_i+m_j\bsl{r}_j}{m_i+m_j}-\bsl{r}_k\right)\end{equation}
($i,j,k$ cyclic), where $\mu_{ij}$ and $\mu_{ij,k}$ are the
appropriate reduced masses
\begin{equation}\mu_{ij}=\frac{m_im_j}{m_i+m_j},~~
\mu_{ij,k}=\frac{(m_i+m_j)m_k}{m_i+m_j+m_k},\end{equation} and
$\mu$ is an arbitrary parameter with the dimension of mass which
drops off in the final expressions. The coordinate
$\bsl{\rho}_{ij}$ is proportional to the separation of quarks $i$
and $j$ and coordinate $\bsl{\lambda}_{ij}$ is proportional to the
separation of quarks $i$ and $j$, and quark $k$. There are three
equivalent ways of introducing the Jacobi coordinates, which are
related to each other by linear transformations with the
coefficients depending on quark masses, with Jacobian equal to
unity. In what follows we omit indices $i,j$.

In terms of the Jacobi coordinates the kinetic energy operator
$H_0$ is written as
\begin{equation} \label{H_0_jacobi} H_0= -\frac{1}{2\mu}
\left(\frac{\partial^2}{\partial\bsl{\rho}^2}
+\frac{\partial^2}{\partial\bsl{\lambda}^2}\right)
=-\frac{1}{2\mu}\left( \frac{\partial^2}{\partial
R^2}+\frac{5}{R}\frac{\partial}{\partial R}+
\frac{K^2(\Omega)}{R^2}\right), \end{equation} where $R$ is the
six-dimensional hyperradius,
\begin{equation} R^2=\bsl{\rho}^2+\bsl{\lambda}^2,\end{equation}
$\Omega$ denotes five residuary angular coordinates, and
$K^2(\Omega)$ is an angular operator whose eigenfunctions (the
hyperspherical harmonics) are
\begin{equation}
\label{eigenfunctions} K^2(\Omega)Y_{[K]}=-K(K+4)Y_{[K]},
\end{equation}
with $K$ being the grand orbital momentum. In terms of $Y_{[K]}$
the wave function $\psi(\bsl{\rho},\bsl{\lambda})$ can be written
in a symbolical shorthand as
\[\psi(\bsl{\rho},\bsl{\lambda})=\sum\limits_K\psi_K(R)Y_{[K]}(\Omega).\]

In the hyperradial approximation, which we shall use below, $K=0$
and $\psi=\psi(R)$. Obviously, this  wave function is completely
symmetric under quark permutations. Note that the centrifugal
potential in the Schr\"odinger equation for the reduced radial
function $\chi(R)=R^{5/2}\psi_K(R)$ with a given $K$ \[
\frac{(K+2)^2-1/4}{R^2}\] is not zero even for $K=0$.

The Coulomb potential is  directly expressed in terms of Jacobi
coordinates:
\begin{equation}
V_C=-\frac{2\alpha_s}{3}\sum\limits_{i<j}\sqrt{\frac{\mu_{ij}}{\mu}}\cdot\frac{1}{|\bsl{\rho}_{ij}|},
\label{Coulomb_Jacobi}
\end{equation}
while the corresponding expression for the string potential is
more complicated. We will  construct it in the next section.

\subsection{String junction point}

Consider the definition of the minimal length string Y--shaped
configuration. Let $\varphi_{ijk}$ be the angle between the line
from quark $i$ to quark $j$ and that from quark $j$ to quark $k$.
One should distinguish two cases. If all $\varphi_{ijk}$ are
smaller than $120^\circ$, and the equilibrium junction position
$\bsl{X}$ coincides with the so-called Torrichelli point of the
triangle in which vertices three quarks are situated. If
$\varphi_{ijk}$ is equal to or greater than $120^\circ$, the
lowest energy configuration has the junction at the position of
quark $j$.

To find the Torrichelli point, consider a scalar function
$\mathcal{L}(\bsl{r})$ of a point inside a triangle
$\triangle\mathrm{ABC}$, defined as a sum of distances between
this point and the triangle vertices:
\begin{equation}
\mathcal{L}(\bsl{r})=|\bsl{r}-\bsl{r}_A|+|\bsl{r}-\bsl{r}_B|+|\bsl{r}-\bsl{r}_C|.
\label{SJ_1}
\end{equation}
The position of the minimum of the function $\mathcal{L}$ is
calculated from the condition
$\dfrac{d\mathcal{L}}{d\bsl{r}}=\bsl{0}$, i.e.:
\begin{equation}
\frac{\bsl{r}-\bsl{r}_A}{|\bsl{r}-\bsl{r}_A|}+
\frac{\bsl{r}-\bsl{r}_B}{|\bsl{r}-\bsl{r}_B|}+
\frac{\bsl{r}-\bsl{r}_C}{|\bsl{r}-\bsl{r}_C|}=-\bsl{n}_A-\bsl{n}_B-\bsl{n}_C=\bsl{0},
\label{SJ_2}
\end{equation}
where $\bsl{n}_{A,B,C}$ are the unit vectors from the Torrichelli
point directed to the vertices of the triangle. It follows from
equation (\ref{SJ_2}) that this condition can be realized only in
the case when angles between vectors $\bsl{n}_A$, $\bsl{n}_B$, and
$\bsl{n}_C$ are equal to $120^\circ$. If $\varphi_{ijk}$ are all
smaller than $120^\circ$, this point exists and is the unique one.
From this point all sides of the triangle are seen at angle of
$120^\circ$.

The geometrical construction of the Torrichelli point is presented
on Fig. 2. One should plot three equilateral triangles
$\triangle\mathrm{AFB}$, $\triangle\mathrm{BDC}$,
$\triangle\mathrm{CEA}$ on the sides of the initial triangle
$\triangle\mathrm{ABC}$. It is lightly to prove the following
statements:
\begin{itemize}
\item The straight lines $\mathrm{AD}$, $\mathrm{BE}$, $\mathrm{CF}$ are crossed at a unique point $\mathrm{T}$;
\item $\mathrm{AD}=\mathrm{BE}=\mathrm{CF}=\mathrm{AT}+\mathrm{BT}+\mathrm{CT}$;
\item $\angle\mathrm{ATF}=\angle\mathrm{BTF}=\angle\mathrm{BTD}=\angle\mathrm{CTD}=\angle\mathrm{CTE}=\angle\mathrm{ATE}
=\dfrac{\pi}{3}$.
\end{itemize}
Now one can easy obtain an expression for a radius-vector of the
Torrichelli point in terms of the lengths $l_i$ of the segments
between this point and the $i^{\text{th}}$ quark (segments
$\mathrm{AT}$, $\mathrm{BT}$, $\mathrm{CT}$ on Fig. 2), and the
quark positions $\bsl{r}_i$ (points $\mathrm{A}$, $\mathrm{B}$,
$\mathrm{C}$ on Fig. 2) \cite{CP02} :
\begin{equation}
\bsl{X}=\frac{l_2l_3\bsl{r}_1+l_1l_3\bsl{r}_2+l_1l_2\bsl{r}_3}{l_2l_3+l_1l_3+l_1l_2}.
\end{equation}
An equivalent expression for $\bsl{X}$ in terms of the
center-of-mass position $\rcm$, and vectors $\bsl{\rho}$ and
$\bsl{\lambda}$ is \cite{NT02}
\begin{equation} \bsl{X}=\rcm+\alpha\bsl{\rho}+\beta\bsl{\lambda}, \label{torr}
\end{equation} with
\[
\begin{aligned}
\alpha&=\frac{1}{2}\sqrt{\frac{\mu}{\mu_{ij}}}\left(\frac{m_j-m_i}{m_i+m_j}-
\frac{1}{\sqrt{3}}\cdot \frac{4t+(3-t^2)\cot\chi}{1+t^2}\right),\\
\beta&=\frac{\sqrt{\mu\mu_{ij,k}}}{m_i+m_j}+
\sqrt{\frac{\mu}{3\mu_{ij}}}\cdot
\frac{\rho}{2\lambda\sin\chi}\cdot\frac{3-t^2}{1+t^2},
\end{aligned}
\]
where
\[
t=\frac{2\lambda\sin\chi+\sqrt{\dfrac{3\mu_{ij,k}}{\mu_{ij}}}\rho}
{2\lambda\cos\chi+\sqrt{\dfrac{3\mu_{ij,k}}{\mu_{ij}}}\cdot\dfrac{m_j-m_i}{m_i+m_j}\rho},
\]
and $\chi$ is the angle between $\bsl{\rho}$ and $\bsl{\lambda}$,
$0\leq\chi\leq \pi$ . It can be easily seen that dependence on
$m_i$ in Eq. (\ref{torr}) is apparent and $\bsl{X}$ does not
depend on quark masses just as it should be.

Eq. (\ref{torr}) is not very useful for our purposes. What we
really need is  the explicit expression for $l_{\text{min}}^2$ in
terms of the Jacobi coordinates \cite{NPV02}. Introducing the
variable $\theta=\arctan(\rho/\lambda)$, $0\leq\theta\leq \pi/2$
one obtains for the case $\varphi_{ijk}<120^\circ$:
\begin{multline}
l^2_{\text{min}}=\mu R^2\cos^2\theta\times{}\\
{}\times\left(\frac{(m^3_1-m^3_2)\tan^2\theta}
{m_1m_2(m^2_1-m^2_2)}+ \left(\frac{m_2-m_1}{m_2+m_1}\cos\chi
+\sqrt{3}\sin\chi\right)\frac{\tan\theta}{m}+\frac{1}{\mu_{12,3}}\right),
\label{l2}
\end{multline}
where $m^2=m_1m_2m_3/(m_1+m_2+m_3)$. If $m_1=m_2=m_3$, this
expression coincides with that derived in \cite{FS91}.

If $\varphi_{ijk}>120^\circ$, $ l_{\text{min}}=r_{ij}+r_{jk}$,
where
\begin{equation}
\begin{aligned}
r_{12}=&R\sin\theta\cdot\sqrt{\frac{\mu}{\mu_{12,3}}}\\
r_{13}=&R\cos\theta\cdot\sqrt{\frac{\mu}{\mu_{12,3}}}\cdot
\sqrt{\frac{m^2}{m_1^2}\tan^2\theta+\frac{2m}{m_1}\tan\theta\cos\chi+1}\\
r_{23}=&R\cos\theta\cdot\sqrt{\frac{\mu}{\mu_{12,3}}}\cdot
\sqrt{\frac{m^2}{m_2^2}\tan^2\theta-\frac{2m}{m_2}\tan\theta\cos\chi+1}
\end{aligned}
\label{r_123}
\end{equation}

The boundaries corresponding to the condition
$\varphi_{ijk}=120^\circ$ in the $(\chi,\theta)$ plane are:
\begin{equation}
\begin{aligned}
\theta_{1(2)}(\chi)&=\arctan(m_{1(2)}(\mp\cos\chi-\sin\chi/\sqrt3)/m),\\
\theta_3(\chi)&=\arctan(m_2(f(\chi)+\sqrt{f^2(\chi)+4\varkappa})/2m),
\end{aligned}
\label{theta_123}
\end{equation}
where
\begin{equation}
f(\chi)=(1-\varkappa)\cos\chi+(1+\varkappa)\sin\chi/\sqrt{3},
\end{equation}
and $\varkappa=m_1/m_2$. These boundaries are shown in Fig. 3 for
the case of equal quark masses.

For simplicity,  the string junction point is often chosen as
coinciding with the center-of-mass coordinate.  In this case
\begin{equation}
\label{cma} V_{\text{string}}=\sigma\sum\limits_{(i,j,k)}
\frac{1}{m_k}\cdot\sqrt{\mu\mu_{ij,k}}\cdot
|\bsl{\lambda}_{ij}|\end{equation}  ($i,j,k$ cyclic). Accuracy of
this approximation that greatly simplifies the calculations was
discussed in \cite{FS91}, \cite{NPV02}. We shall comment on this
point later on.

\subsection{Hyperradial approximation}

Introducing the variable $x=\sqrt{\mu} R$ and averaging the
interaction $U=V_C+ V_{\text{string}}$ over the six-dimensional
sphere $\Omega_6$, one obtains the Schr\"odinger equation for
$\chi(x)$:
\begin{equation} \label{shr}
\frac{d^2\chi(x)}{dx^2}+2\left[E_0+\frac{a}{x}-bx-\frac{15}{8x^2}\right]\chi(x)=0.
\end{equation}
Because the wave function $\psi$ must be finite at the origin
$\chi(x) \sim {\cal O} (x^{5/2})$ as $x\to 0$. As $x\to \infty$
one can neglect the Coulomb-like and centrifugal terms, and Eq.
(\ref{shr}) becomes
\begin{equation}
\frac{d^2\chi(z)}{dz^2}-z\chi(z)=0,\quad z=(2b)^{1/3}x.
\end{equation}
This is the familiar Airy equation whose solution $\mathrm{Ai}(z)$
behaves at infinity  as
\begin{equation} \mathrm{Ai}(z)\sim\frac{1}{2}\pi^{-1/2}z^{-1/4}\exp(-\frac{2}{3}z^{3/2}),
~~ \mathrm{Re}~z\geq~0. \end{equation} In Eq. (\ref{shr}) $E_0$ is
the ground-state eigenvalue and
\begin{equation}
\begin{aligned}
a&=R\sqrt{\mu}\cdot\int V_C(\bsl{r}_1,\bsl{r}_2,\bsl{r}_3)\cdot
d\Omega_6~=~\frac{2\alpha_s}{3}\cdot \frac{16}{3\pi}\cdot \left(
\sum\limits_{i<j}\sqrt{\mu_{ij}}\right),\\
b&=\frac{1}{R\sqrt{\mu}}\cdot\int
V_{\rm{string}}(\bsl{r}_1,\bsl{r}_2,\bsl{r}_3)\cdot d\Omega_6.
\end{aligned} \label{ab_int}
\end{equation}

The expression for the coefficient $b$ can not be obtained
analytically except for the equal quark mass system
$m_1=m_2=m_3=m$, in which case the straightforward calculation
yields:
\begin{equation}
b=\frac{4}{15}\cdot\frac{\sigma}{\sqrt{m}}\cdot\left(
-\sqrt{3}+\frac{12\sqrt{2}}{\pi}+\frac{3\sqrt{3}}{\pi}\arccos\frac{1}{5}
\right)\approx 1.58\cdot\frac{\sigma}{\sqrt{m}} . \label{b_exact}
\end{equation}
On the contrary, in the approximation (\ref{cma}) the coefficient
$b$ can be easily found analytically for the case of arbitrary
quark masses:
\begin{equation}\label{b_approx}
b=\sigma\cdot\frac{32}{15\pi}\cdot\left(\sum\limits_{i<j}\frac{\sqrt{\mu_{ij,k}}}{m_k}\right).
\end{equation}

Let us explain the numerical coefficients in (\ref{ab_int}),
(\ref{b_approx})  in more details. To this end we introduce the
angle $\theta$ as in (\ref{l2}), such that
\begin{equation}
\rho=R\sin\theta,~~\lambda=R\cos\theta,~~
0\le\theta\le\frac{\pi}{2},
\end{equation}
and  write the volume element $d^3\rho d^3\lambda$ as
\begin{equation}
\label{phs} d^3\rho~ d^3\lambda=(4\pi)^2\rho^2\lambda^2 d\rho
d\lambda=(4\pi)^2 R^5\sin^2\theta\cos^2\theta dR d\theta.
\end{equation}
The volume of the six-dimensional sphere is
\begin{equation}
\Omega_6=(4\pi)^2\int\limits_0^{\frac{\pi}{2}}\sin^2\theta\cos^2\theta
d\theta =(4\pi)^2\cdot\frac{\pi}{16}=\pi^3.
\end{equation}
Then averaging the Coulomb and string terms yields
\begin{equation}
<\frac{1}{\rho}>=\frac{1}{\pi^3}\cdot\frac{1}{R}\cdot(4\pi)^2\cdot
\int\limits_0^{\frac{\pi}{2}}\sin\theta\cos^2\theta~d\theta=
\frac{16}{3\pi}\cdot\frac{1}{R},\label{rho_av}
\end{equation}
\begin{equation}
<\lambda>=\frac{1}{\pi^3}\cdot R\cdot(4\pi)^2\cdot
\int\limits_0^{\frac{\pi}{2}}\sin^2\theta \cos^3
\theta~d\theta=\frac{32}{15\pi}\cdot R. \label{lambda_av}
\end{equation} Combining together expressions
(\ref{Coulomb_Jacobi}), (\ref{cma}), (\ref{ab_int}),
(\ref{rho_av}), (\ref{lambda_av}) lead to (\ref{ab_int}),
(\ref{b_approx}).

\subsection{Analytic results for light baryons}

We can eliminate all dimensional parameters from the equation
(\ref{shr}) by a substitution $y=b^{1/3}x$, which leads us to the
equation:
\begin{equation}
\frac{d^2\chi}{dy^2}+2\left(\mathcal{E}-y+\frac{\delta}{y}-\frac{15}{8y^2}\right)=0
\label{an_shr}
\end{equation}
where
\[
\mathcal{E}=E_0 b^{-2/3},\quad \delta=ab^{-1/3}.
\]

The eigenvalue of the Eq. (\ref{an_shr}) can be found using the
ordinary perturbation theory, considering the Coulomb term
$\left(-\dfrac{\delta}{y}\right)$ as a small perturbation. This
approximation works well for a nucleon containing three light
quarks with the running mass equal to zero. In this case there is
only one dynamical quark mass $m$. So, the task is greatly
simplified and one can obtain analytic expressions for $m$ and
$M_B$ via two parameters: $\sigma$ and $\alpha_s$, as expansions
in powers of $\alpha_s$.

Omitting the intermediate details outlined in Appendix we quote
here the result with accuracy up to $\alpha_s^2$:
\begin{equation}
\label{m_q}
m=
0.959\cdot\sqrt{\sigma}\cdot(1+0.270\alpha_s+0.117\alpha_s^2+\dots),
\end{equation}
\begin{equation}
\label{m_baryon}M_B=
5.751\cdot\sqrt{\sigma}\cdot(1-0.270\alpha_s-0.039\alpha_s^2+\dots)+C,
\end{equation} or
\begin{equation}
M_B=6m\cdot(1-0.540\alpha_s-0.083\alpha_s^2+\dots)+C,\label{m_b}
\end{equation} where the constant $C$ has been defined in Eq. (\ref{self_energy}).
As it follows from Eq. (\ref{m_b}), the Coulomb-like correction to
$M_B$ comprises approximately $20\%$.

\subsection{Quasi-classical solution}

For the purpose of illustration, the problem is first solved
quasi-classically rather than using quantum mechanics. This
approach is based on the well known fact that interplay between
the centrifugal term and the confining potential produces an
effective potential minimum  specific for the three-body problem.
The numerical solution of (\ref{shr}) for the ground-state
eigenenergy may be reproduced on a percent level of accuracy by
using the parabolic approximation \cite{KNS87} for the effective
potential
\[ U(x)=-\frac{a}{x}+bx+\frac{15}{8x^2}. \]
This approximation provides an analytical expression for the
eigenenergy. The potential $U(x)$ has the minimum at a point
$x=x_0$, which is defined by the condition $ U'(x_0)=0$, i.e.:
\begin{equation} \label{x_0} b
x_0^3+ax_0-15/4=0\, . \end{equation} Expanding $U(x)$ in the
vicinity of the minimum one obtains \[ U(x)\approx
U(x_0)+\frac{1}{2}U''(x_0)(x-x_0)^2, \] i.e., the potential of the
harmonic oscillator with the frequency $\omega=\sqrt{ U''(x_0)}$.
Therefore the ground-state energy eigenvalue is
\begin{equation} \label{eigenvalue} E_0\approx
U(x_0)+\frac{1}{2}\omega\, .
\end{equation}

\subsection{Variational solution}

Another rather accurate method of solving Eq. (\ref{shr}) is the
minimization of the baryon energy using  a simple variational
Ans\"atz
\begin{equation}\label{gaussian}\chi(x)\sim x^{5/2}e^{-
p^2x^2},\end{equation} where $p$ is the variational parameter.
Then using the three-quark Hamiltonian one can get an approximate
expression for the ground-state energy: $E_0\approx\min\limits_p
E_0(p)$, where
\begin{equation}\ \label{E_0_var}
E_0(p)=\langle\chi|H|\chi\rangle=
3p^2-a\cdot\frac{3}{4}\cdot\sqrt{\frac{\pi}{2}}\cdot
p+b\cdot\frac{15}{16}\cdot\sqrt{\frac{\pi}{2}}\cdot\frac{1}{ p}.
\end{equation}
The accuracy of the variational solution will be illustrate in
Sect. 4.

\subsection{Analytical results for $(Qud)$ baryons}

For the heavy quarks (Q=c and b) the variation in the values of
their masses $m_Q$ is marginal. This is illustrated by the simple
analytical results for $(Qud)$ baryons \cite{NT02}. These results
were obtained from the approximate solution of equation
\begin{equation} \frac{\partial E_0(m_1,m_2,m_3,p)}{\partial p}=0\end{equation} where $E_0$ is given by Eq. (\ref{E_0_var})
in the form of expansion in the small parameters
\begin{equation}\xi=\frac{\sqrt{\sigma}}{m_Q^{(0)}}~~ {\rm and}~~ \alpha_s,\end{equation}
where $m_Q^{(0)}$ is the heavy-quark running mass.

Omitting the intermediate steps one obtains:
\begin{eqnarray}
E_0&=&3\sqrt{\sigma}\left(\frac{6}{\pi}\right)^{1/4}\left(1+A\cdot\xi
-\frac{5}{3}B\cdot\alpha_s+\dots\right),\\
m_q&=&\sqrt{\sigma}\left(\frac{6}{\pi}\right)^{1/4}\left(1-A\cdot\xi+B\cdot\alpha_s+
\dots\right),\\ \label{m_Q} m_Q&=&m_Q^{(0)}\left(1+{\cal
O}(\xi^2,\alpha_s^2, \alpha_s\xi)+\dots\right), \end{eqnarray}
where for the Gaussian variational Ans\"atz
(\ref{gaussian})\begin{equation}
A=\frac{\sqrt{2}-1}{2}\left(\frac{6}{\pi}\right)^{1/4}\approx
0.24,~ B=\frac{4+\sqrt{2}}{18}\sqrt{\frac{6}{\pi}}\approx
0.42.\end{equation} Note that the corrections of the first order
in $\xi$ and $\alpha_s$ are absent in the expression (\ref{m_Q})
for $m_Q$. Accuracy of this approximation is illustrated in Table
1 of  \cite{NT02}.

\section{Baryon masses}

\subsection{Quark dynamical masses}

We first calculate the  dynamical masses $m_i$ retaining only the
string potential in the effective Hamiltonian (\ref{EH}). This
procedure is in agreement with the strategy adopted in
\cite{KS00}. The masses $m_i$ are then obtained from the equation:
\begin{equation}
\frac{\partial M_B^{(0)}}{\partial m_i}=0,
\end{equation}
where \begin{equation}
M_B^{(0)}=\sum\limits_{i=1}^3\left(\frac{{m_i^{(0)}}^2}{2m_i}+\frac{m_i}{2}\right)+E_0(m_1,m_2,m_3;\alpha_s=0).
\end{equation}
Then we add the one-gluon-exchange Coulomb potential and solve Eq.
(\ref{shr}) to obtain the ground-state eigenvalues
$E_0(m_1,m_2,m_3;\alpha_s)$ for a given $\alpha_s$. The physical
mass $M_B$ of a baryon is
\begin{equation}\label{M_B_full}
M_B=\sum\limits_{i=1}^3\left(\frac{{m_i^{(0)}}^2}{2m_i}+\frac{m_i}{2}\right)+E_0(m_1,m_2,m_3;\alpha_s)+C.
\end{equation}

We use the values of parameters $\sigma=0.15\text{~GeV}^2$ (this
value has been found in a recent lattice study \cite{TMNS02}) ,
$\alpha_s=0.39$, $m^{(0)}_q=0.009\text{~GeV}$ ($q=u,d$),
$m^{(0)}_s=0.17\text{~GeV}$, $m^{(0)}_c=1.4\text{~GeV}$, and
$m^{(0)}_b=4.8\text{~GeV}$.
 The results for various
baryons, obtained using various approximations, are given in
Tables 1--3. Table 1 contains the results obtained using the
variational solution of Eq. (\ref{shr}) using  the approximation
(\ref{cma}) for the three-quark potential . In Table 2 the results
are shown obtained using the same approximation, $\bsl{X}=\rcm$,
but exact numerical solution of Eq. (\ref{shr}). Table 3 contains
the results obtained by the numerical integration of (\ref{shr})
with the genuine three-quark potential in the form (\ref{l2}),
(\ref{r_123}). Comparing the results of Tables 1 and 2 we observe
a good accuracy of the variational solution: the difference
between variational and exact results for $M_B$ does not exceed
$10-15\text{~MeV}$ for all baryons from lightest to doubly heavy
ones. The approximation $\bsl{X}=\rcm$ leads to a $\sim 5\%$
increase of the coefficient $b$ in (\ref{shr}). As a consequence,
this approximation increases the baryon masses by $\sim
70\text{~MeV}$ (compare the results of Tables 2 and 3.

Note that there is no good theoretical reason why quark masses
$m_i$ need to be the same in different baryons. Inspection of
Table 1 shows that the masses of the light quarks ($u$, $d$, or
$s$) are increased by $\sim 100\text{~MeV}$ when going from light
to heavy baryons. The dynamical masses of light quarks
$m_{u,d,s}\sim\sqrt{\sigma}\sim~400-500\text{~MeV}$ qualitatively
agree with the results of \cite{KN00} obtained from the analysis
of the heavy-light ground-state mesons.

While studying Table 3 is sufficient to have an appreciation of
the accuracy of our predictions, few comments should be added. We
expect an accuracy of the baryon predictions to be $\sim 5-10\%$
that is partly due to the approximations employed in the
derivations of the Effective Hamiltonian itself \cite{Si02} and
partly due to the error associated with the variational nature of
hyperspherical approximation. From this point of view the overall
agreement with data is quite satisfactory. E.g. we get
$\dfrac{1}{2}(N+\Delta)_{\text{theory}}=1144\text{~MeV}$ vs.
$\dfrac{1}{2}(N+\Delta)_{\text{exp}}=1085\text{~MeV}$ ( a $5\%$
increase in $\alpha_s$ would correctly give the $(N,\Delta)$
center of gravity),
$\dfrac{1}{4}(\Lambda+\Sigma+2\Sigma^*)=1242\text{~MeV}$ vs.
experimental value of $1267\text{~MeV}$. We also find
$\Xi_{\text{theory}}=1336\text{~MeV}$ (without hyperfine
splitting) vs. $\Xi_{\text{exp}}=1315\text{~MeV}$ and
$\Xi^c_{\text{theory}}=2542\text{~MeV}$ vs.
$\Xi^c_{\text{exp}}=2584\text{~MeV}$. On the other hand, our study
shows some difficulties in reproducing the $\Omega$-hyperon mass.

\subsection{Doubly heavy baryons}

In Table 4 we compare the spin-averaged masses (computed without
the spin-spin term) of the lowest doubly heavy baryons to the
predictions of other models \cite{GKLO00}, \cite{EFGM02},
\cite{BDGNR94} as well as variational calculations of \cite{NT01}
for which the center-of-gravity of non-strange baryons and
hyperons was essential a free parameter. Most of recent
predictions were obtained in a light quark--heavy diquark model
\cite{GKLO00}, \cite{EFGM02}, in which case the spin-averaged
values are $M=\frac{1}{3}(M_{1/2}+2M_{3/2})$. Note that the wave
function calculated in the hyperspherical approximation shows the
marginal diquark clustering in the doubly heavy baryons. This is
principally kinematic effect related to the fact that in this
approximation the difference between the various mean values
$\bar{r}_{ij}$ in a baryon is due to the factor
$\sqrt{1/\mu_{ij}}$ which varies between $\sqrt{2/m_i}$ for
$m_i=m_j$ and $\sqrt{1/m_i}$ for $m_i\ll m_j$. In general, in
spite of the completely different physical picture, we find a
reasonable agreement within $100\text{~MeV}$ between different
predictions for the ground-state masses of the doubly heavy
baryons. Our prediction for $M_{ccu}$ is $3.66\text{~GeV}$ with
the perturbative hyperfine splitting $\Xi^*_{ccu}-\Xi_{ccu}\sim 40
\text{~MeV}$. This splitting has been calculated using the
Fermi--Breit spin-spin interaction \cite{deRuhula}. The change of
$\sigma$ to $0.17\text{~GeV}^2$ increases the mass of $\Xi_{cc}$
by $\sim 30\text{~MeV}$. Note that the mass of $\Xi_{cc}$ is
rather sensitive to the value of the running $c$-quark mass
$m_c^{(0)}$, see Fig. 4.

\section{Conclusions}

In this paper, we have outlined a novel approach to baryon
spectroscopy which is based on a single framework of the Effective
Hamiltonian that is consistent with QCD. This model uses the
stringlike picture of confinement and perturbative
one-gluon-exchange potential. The main advantage of this work is
demonstration of the fact that it is possible to describe all the
baryons in terms of the only two parameters inherent to QCD,
namely $\sigma$ and $\alpha_s$.

\section*{Acknowledgements}

This work was supported in part by NATO grant \# PST.CLG.978710
and by RFBR grants \#\# 03-02-17345, 02-02-17379.

\section*{Appendix}



\begin{center}
{\bfseries The approximate calculation of the mass of the
three-light-quark system in the hyperspherical formalism}
\end{center}

Let us consider the case when each quark has a zero current mass
and the constituent mass $m$, which is the same for all three
quarks. Then the mass of the system is the minimum of the function
$\mathcal{H}(m)$: $\mathcal{M}=\min\limits_m \mathcal{H}(m)$,
where\footnote{For simplicity we omit here and below the
corrections $C$.}
\begin{equation}
\mathcal{H}(m)=\frac{3m}{2}+E\, , \label{A_H}
\end{equation}
and $E$ is the energy level defined from the ordinary Shr\"odinger
equation (see Eq. (\ref{shr})):
\begin{equation}
\frac{d^2\chi}{dx^2}+2\left(E+
\frac{\beta\sqrt{m}}{x}-\frac{\gamma}{\sqrt{m}}x-\frac{15}{8x^2}\right)\chi=0\,
, \label{A_Sh}
\end{equation}
where
\begin{eqnarray}
\beta&=&\frac{16\sqrt{2}}{3\pi}\alpha_s\approx 2.401\cdot\alpha_s,
\label{A_beta}\\ \gamma&=&\frac{32\sqrt{6}}{15\pi}\sigma\approx
1.663\cdot\sigma \, .\label{A_gamma}
\end{eqnarray}

Let us replace a variable $x$ by a dimensionless variable $y$:
$y=\left(\dfrac{\gamma^2}{m}\right)^{1/6}x$. Then:
\begin{equation}
\frac{d^2\chi}{dy^2}+2\left(\mathcal{E}+\frac{\delta}{y}-y-\frac{15}{8y^2}\right)\chi=0\,
, \label{A_Sh1}
\end{equation}
where $\mathcal{E}$ and $\delta$ are the dimensionless parameters:
\[
\mathcal{E}=E\left(\frac{m}{\gamma^2}\right)^{1/3},\quad
\delta=\beta\left(\frac{m^2}{\gamma}\right)^{1/3}.
\]

We will calculate the eigenvalues of the equation (\ref{A_Sh1})
using the second-order perturbation theory, considering the
Coulomb term $\left(-\dfrac{\delta}{y}\right)$ in the potential as
a small perturbation. The unperturbed equation is:
\begin{equation}
\frac{d^2f}{dy^2}+2\left(\lambda-y-\frac{15}{8y^2}\right)f=0\, .
\label{A_MyEq}
\end{equation}
It contains no physical parameters, so its solution is a pure
mathematical task. Let us denote the eigenvalues of the equation
(\ref{A_MyEq}) as $\{\lambda_i\}$:
\[
0<\lambda_0<\lambda_1<\dots\, ,
\]
and the corresponding normalized eigenfunctions as $\{f_i(y)\}$.
In what follows we will use the notations:
\begin{eqnarray*}
\xi&=&\int\limits_0^{+\infty}\frac{1}{y}f_0^2(y)dy\, ,\quad
\xi>0,\\ \eta&=&\sum\limits_{i=1}^\infty
\frac{\left(\int\limits_0^{+\infty}\dfrac{1}{y}f_i(y)f_0(y)dy\right)^2}{\lambda_i-\lambda_0}\,
,\quad \eta>0.
\end{eqnarray*}
The approximate numerical values of these parameters are:
\begin{eqnarray}
\lambda_0&\approx&3.030, \label{A_lambda} \\ \xi&\approx&0.553,
\label{A_xi} \\ \eta&\approx&0.028. \label{A_eta}
\end{eqnarray}

The ground level of the equation (\ref{A_Sh1}) can be
approximately calculated as follows:
\begin{equation}
\mathcal{E}\approx\lambda_0-\delta\xi-\delta^2\eta\, .
\label{A_Energy}
\end{equation}
The small parameter here is the ratio
$\dfrac{\delta\xi}{\lambda_0}$. To estimate this quantity one can
solve the task in the zero approximation without the Coulomb term.
So,
\begin{eqnarray*}
\mathcal{E}&\approx&\lambda_0,\\
E&\approx&\lambda_0\left(\frac{\gamma^2}{m}\right)^{1/3},\\
\mathcal{H}&\approx&\frac{3}{2}m+\lambda_0\left(\frac{\gamma^2}{m}\right)^{1/3}.
\end{eqnarray*}
$\mathcal{H}$ has a single minimum, defined by an equation
$\partial\mathcal{H}/\partial m=0$, i.e.,
\[
\frac{3}{2}-\frac{1}{3}\lambda_0\gamma^{2/3}m^{-4/3}=0.
\]
Thus $m=\left(\dfrac{2\lambda_0}{9}\right)^{3/4}\sqrt{\gamma}$,
and finally:
\[
\frac{\delta\xi}{\lambda_0}=\frac{\beta\xi}{\lambda_0}\left(\frac{m^2}{\gamma}\right)^{1/3}=
\frac{2\beta\xi}{3\sqrt{2\lambda_0}}\approx 0.360\alpha_s\, .
\]
For $\alpha_s=0.4$ one has $\dfrac{\delta\xi}{\lambda_0}\approx
0.14\ll 1$. This verifies the correctness of using the
perturbation theory in this task.

Now, using Eq. (\ref{A_Energy}), we can calculate the constituent
mass $m$ and the mass of the state $\mathcal{M}$. For the energy
level we have:
\[
E=\left(\frac{\gamma^2}{m}\right)^{1/3}\mathcal{E}=\left(\frac{\gamma^2}{m}\right)^{1/3}\lambda_0
-(\gamma m)^{1/3}\beta\xi-m\beta^2\eta\, .
\]

It is convenient to use the positive dimensionless parameter
$q=m^{1/3}\gamma^{-1/6}$, so that $m=\sqrt{\gamma}q^3$. Then
\[
E=\sqrt{\gamma}(\lambda_0 q^{-1}-\beta\xi q-\beta^2\eta q^3)\, ,
\]
and substituting it to the Eq. (\ref{A_H}), we find:
\[
\mathcal{H}=\sqrt{\gamma}\left(\lambda_0 q^{-1}-\beta\xi
q+\left(\frac{3}{2}-\beta^2\eta\right)q^3\right)\, .
\]

$\mathcal{H}$ has a single minimum, defined from the condition:
$\left.\dfrac{\partial\mathcal{H}}{\partial q}\right|_{q=q_0}=0$,
i.e.,
\begin{equation}
\left(\frac{9}{2}-3\beta^2\eta\right)q_0^2-\beta\xi-\frac{\lambda_0}{q_0^2}=0\,
. \label{A_Min}
\end{equation}
After calculating $q_0$, one can find the constituent mass
$m=\sqrt{\gamma}q_0^3$ and the mass of the system $\mathcal{M}$:
\[
\mathcal{M}=\mathcal{H}(q_0)=\frac{2\sqrt{\gamma}}{3}\left(\frac{2\lambda_0}{q_0}-\beta\xi
q_0 \right)\, .
\]

The equation (\ref{A_Min}) can be lightly solved:
\begin{equation}
q_0=\sqrt{\frac{\beta\xi+\sqrt{\beta^2\xi^2+2\lambda_0(9-6\beta^2\eta)}}{9-6\beta^2\eta}}\,
. \label{A_p0}
\end{equation}
On expanding the right-hand side of this equation taking $\beta$
as a small parameter
\[
q_0\approx\frac{(2\lambda_0)^{1/4}}{\sqrt{3}}\left(1+\frac{\xi\beta}{6\sqrt{2\lambda_0}}+
\left(\frac{\xi^2}{144\lambda_0}+\frac{\eta}{6}\right)\beta^2\right)\,
,
\]
we easily get formulas for the constituent mass:
\begin{equation}
m\approx\sqrt{\gamma}\frac{(2\lambda_0)^{3/4}}{3\sqrt{3}}\left(1+\frac{\xi\beta}{2\sqrt{2\lambda_0}}+
\left(\frac{\xi^2}{16\lambda_0}+\frac{\eta}{2}\right)\beta^2\right)\,
, \label{A_m}
\end{equation}
and for the state mass:
\begin{equation}
\mathcal{M}\approx\sqrt{\gamma}(2\lambda_0)^{3/4}\frac{2}{\sqrt{3}}\left(1-\frac{\xi\beta}{2\sqrt{2\lambda_0}}-
\left(\frac{\xi^2}{48\lambda_0}+\frac{\eta}{6}\right)\beta^2\right)\,
. \label{A_M}
\end{equation}

Substituting in the equations (\ref{A_m}) and (\ref{A_M})
numerical values according to formulas (\ref{A_beta}),
(\ref{A_gamma}), (\ref{A_lambda}), (\ref{A_xi}), (\ref{A_eta}), we
obtain the results quoted in Eqs. (\ref{m_b}, (\ref{m_baryon}).

For $\sigma=0.15\text{~GeV}^2$, $\alpha_s=0.4$ one has $m\approx
0.418\text{~GeV}$, $\mathcal{M}\approx 1.973\text{~GeV}$.

\newpage

\section*{Tables}

\noindent {\bfseries Table 1.} Summary of variational calculations
of Eq. (\ref{shr}) for the various baryon states in the
approximation $\bsl{X}=\rcm$.

{\normalsize \begin{center}
\begin{tabular}{|c|c|c|c|c|c|}
\hline Baryon & $m_1$ & $m_2$ & $m_3$ & $E_0$ & $M_B$
\\ \hline $(qqq)$ & 0.372 & 0.372 & 0.372 & 1.433 & 1.221
\\ \hline $(qqs)$ & 0.377 & 0.377 & 0.415 & 1.404 & 1.314
\\ \hline $(qss)$ & 0.381 & 0.420 & 0.420 & 1.377 & 1.405
\\ \hline $(sss)$ & 0.424 & 0.424 & 0.424 & 1.350 & 1.493
\\ \hline $(qqc)$ & 0.424 & 0.424 & 1.464 & 1.178 & 2.538
\\ \hline $(qsc)$ & 0.427 & 0.465 & 1.467 & 1.153 & 2.613
\\ \hline $(ssc)$ & 0.468 & 0.468 & 1.469 & 1.129 & 2.686
\\ \hline $(qqb)$ & 0.446 & 0.446 & 4.819 & 1.093 & 5.909
\\ \hline $(qsb)$ & 0.448 & 0.487 & 4.820 & 1.067 & 5.978
\\ \hline $(ssb)$ & 0.490 & 0.490 & 4.821 & 1.042 & 6.046
\\ \hline $(qcc)$ & 0.459 & 1.498 & 1.498 & 0.914 & 3.712
\\ \hline $(scc)$ & 0.499 & 1.499 & 1.499 & 0.890 & 3.777
\\ \hline $(qcb)$ & 0.477 & 1.524 & 4.834 & 0.793 & 7.021
\\ \hline $(scb)$ & 0.517 & 1.525 & 4.834 & 0.770 & 7.083
\\ \hline $(qbb)$ & 0.495 & 4.854 & 4.854 & 0.606 & 10.260
\\ \hline $(sbb)$ & 0.534 & 4.855 & 4.855 & 0.583 & 10.318
\\ \hline
\end{tabular}
\end{center}}


\vspace*{10mm}

\noindent {\bfseries Table 2.} The results obtained from the exact
numerical solution of Eq. (\ref{shr}) in the approximation
$\bsl{X}=\rcm$ for the various baryon states .

{\normalsize \begin{center}
\begin{tabular}{|c|c|c|c|c|c|}
\hline Baryon & $m_1$ & $m_2$ & $m_3$ & $E_0$ & $M_B$
\\ \hline $(qqq)$ & 0.372 & 0.372 & 0.372 & 1.427 & 1.212
\\ \hline $(qqs)$ & 0.376 & 0.376 & 0.415 & 1.398 & 1.306
\\ \hline $(qss)$ & 0.381 & 0.419 & 0.419 & 1.370 & 1.397
\\ \hline $(sss)$ & 0.423 & 0.423 & 0.423 & 1.344 & 1.485
\\ \hline $(qqc)$ & 0.424 & 0.424 & 1.464 & 1.170 & 2.530
\\ \hline $(qsc)$ & 0.426 & 0.465 & 1.466 & 1.146 & 2.604
\\ \hline $(ssc)$ & 0.467 & 0.467 & 1.468 & 1.122 & 2.677
\\ \hline $(qqb)$ & 0.445 & 0.445 & 4.820 & 1.085 & 5.900
\\ \hline $(qsb)$ & 0.448 & 0.487 & 4.820 & 1.059 & 5.970
\\ \hline $(ssb)$ & 0.488 & 0.488 & 4.820 & 1.035 & 6.037
\\ \hline $(qcc)$ & 0.458 & 1.497 & 1.497 & 0.905 & 3.702
\\ \hline $(scc)$ & 0.497 & 1.498 & 1.498 & 0.882 & 3.767
\\ \hline $(qcb)$ & 0.475 & 1.523 & 4.833 & 0.784 & 7.010
\\ \hline $(scb)$ & 0.515 & 1.523 & 4.837 & 0.760 & 7.072
\\ \hline $(qbb)$ & 0.490 & 4.850 & 4.850 & 0.596 & 10.245
\\ \hline $(sbb)$ & 0.530 & 4.856 & 4.856 & 0.571 & 10.303
\\ \hline
\end{tabular}
\end{center}}

\pagebreak

\noindent {\bfseries Table 3.} The same as in Table 2 but for an
exact treatment of the string-junction point.

{\normalsize \begin{center}
\begin{tabular}{|c|c|c|c|c|c|}
\hline Baryon & $m_1$ & $m_2$ & $m_3$ & $E_0$ & $M_B$
\\ \hline $(qqq)$ & 0.362 & 0.362 & 0.362 & 1.392 & 1.144
\\ \hline $(qqs)$ & 0.367 & 0.367 & 0.407 & 1.362 & 1.242
\\ \hline $(qss)$ & 0.371 & 0.411 & 0.411 & 1.335 & 1.336
\\ \hline $(sss)$ & 0.415 & 0.415 & 0.415 & 1.307 & 1.426
\\ \hline $(qqc)$ & 0.406 & 0.406 & 1.470 & 1.142 & 2.464
\\ \hline $(qsc)$ & 0.409 & 0.448 & 1.471 & 1.116 & 2.542
\\ \hline $(ssc)$ & 0.452 & 0.452 & 1.473 & 1.090 & 2.621
\\ \hline $(qqb)$ & 0.425 & 0.425 & 4.825 & 1.054 & 5.823
\\ \hline $(qsb)$ & 0.429 & 0.469 & 4.826 & 1.026 & 5.903
\\ \hline $(ssb)$ & 0.471 & 0.471 & 4.826 & 1.000 & 5.975
\\ \hline $(qcc)$ & 0.444 & 1.494 & 1.494 & 0.876 & 3.659
\\ \hline $(scc)$ & 0.485 & 1.496 & 1.496 & 0.851 & 3.726
\\ \hline $(qcb)$ & 0.465 & 1.512 & 4.836 & 0.753 & 6.969
\\ \hline $(scb)$ & 0.505 & 1.514 & 4.837 & 0.729 & 7.032
\\ \hline $(qbb)$ & 0.488 & 4.847 & 4.847 & 0.567 & 10.214
\\ \hline $(sbb)$ & 0.526 & 4.851 & 4.851 & 0.544 & 10.273
\\ \hline
\end{tabular}
\end{center}}

\vspace*{10mm}

\noindent {\bfseries Table 4}. Comparison of various predictions
for ground-state masses (in units of GeV) of doubly heavy baryons.

{\normalsize \begin{center}
\begin{tabular}{|c|c|c|c|c|c|} \hline Baryon & Ref. \cite{NPV02} & Ref. \cite{NT01} &
Ref. \cite{GKLO00} & Ref. \cite{EFGM02} & Ref. \cite{BDGNR94}

\\ \hline $\Xi_{cc}$ & ~3.66 & ~3.69 & ~3.57 & ~3.69 & ~3.70
\\ \hline $\Omega_{cc}$ & ~3.73 & ~3.86 & ~3.66 & ~3.84 & ~3.80
\\ \hline $\Xi_{cb}$ & ~6.97 & ~6.96 & ~6.87 & ~6.96 & ~6.99
\\ \hline $\Omega_{cb}$ & ~7.03 & ~7.13 & ~6.96 & ~7.15 & ~7.07
\\ \hline $\Xi_{bb}$ & 10.21 & 10.16 & 10.12 & 10.23 & 10.24
\\ \hline $\Omega_{bb}$ & 10.27 & 10.34 & 10.19 & 10.38 & 10.34
\\ \hline
\end{tabular}
\end{center}}

\newpage

\section*{Figure captions}

\noindent {\bfseries Figure 1.} Three-lobes Wilson loop.\\

\noindent {\bfseries Figure 2.} The geometrical construction of
the Torrichelli point $\mathrm{T}$ for an arbitrary triangle
$\triangle\mathrm{ABC}$. The triangles $\triangle\mathrm{AFB}$,
$\triangle\mathrm{BDC}$, $\triangle\mathrm{CEA}$ are
equilateral.\\

\noindent {\bfseries Figure 3.} The four regions in the
$(\chi,\theta)$ plane corresponding to $\varphi_{ijk}\ge
120^\circ$ and $\varphi_{ijk} \le 120^\circ$ for the case of equal
quark masses.\\

\noindent {\bfseries Figure 4.} Mass of $\Xi_{cc}$ as a function
of the running $c$-quark mass for $\sigma=0.15\text{~GeV}^2$ and
$\sigma=0.17\text{~GeV}^2$. The masses are given in GeV. Bold
points refer to the case $m_c^{(0)}=1.4\text{~GeV}$.\\

\newpage

\begin{center}
\includegraphics[width=160mm,
keepaspectratio=true]{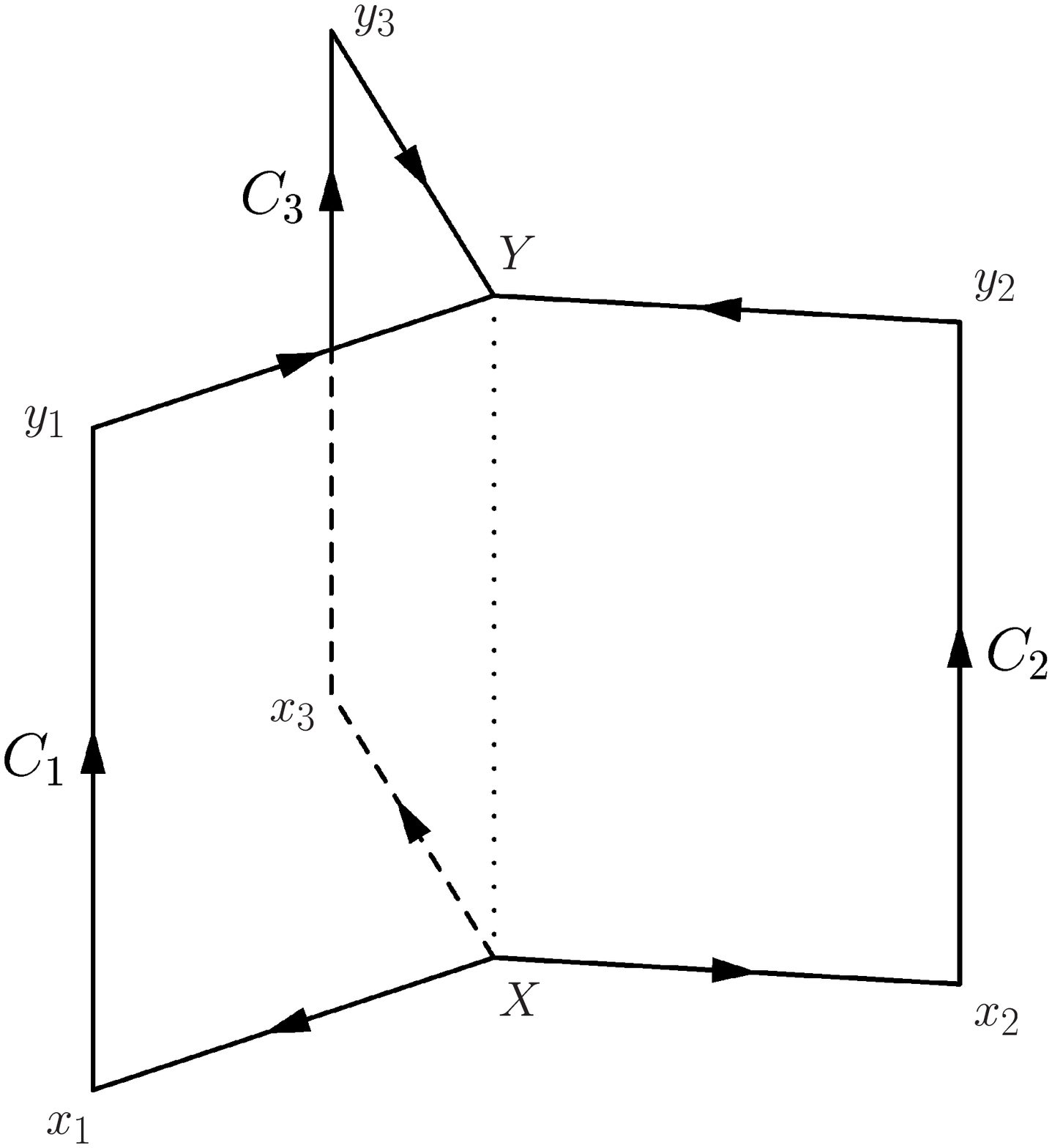}

\vspace*{15mm} {\Large\bfseries Fig.1.}
\end{center}

\newpage

\begin{center}
\includegraphics[width=160mm,
keepaspectratio=true]{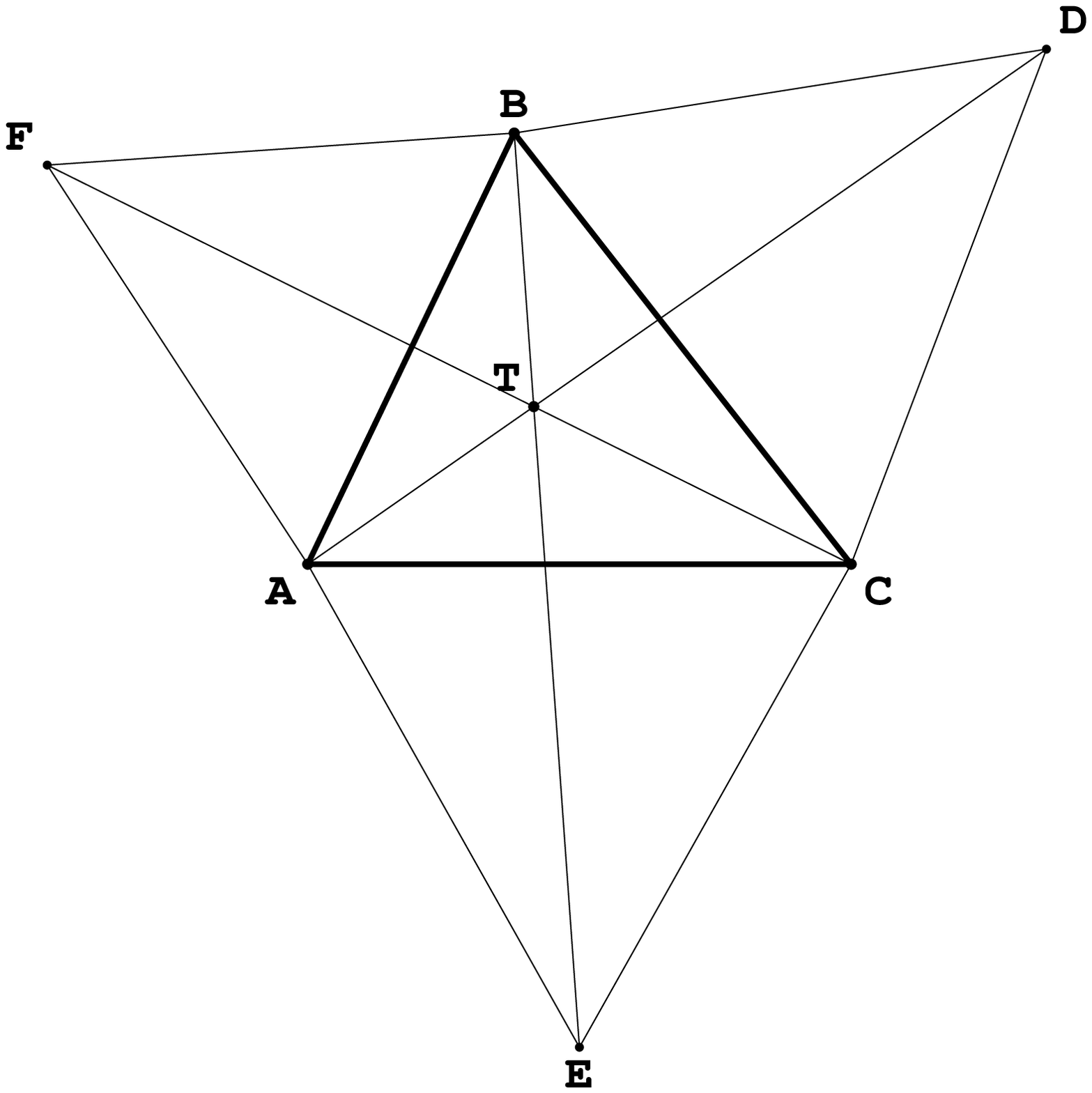}

\vspace*{15mm} {\Large\bfseries Fig.2.}
\end{center}

\newpage

\begin{center}
\includegraphics[width=160mm,
keepaspectratio=true]{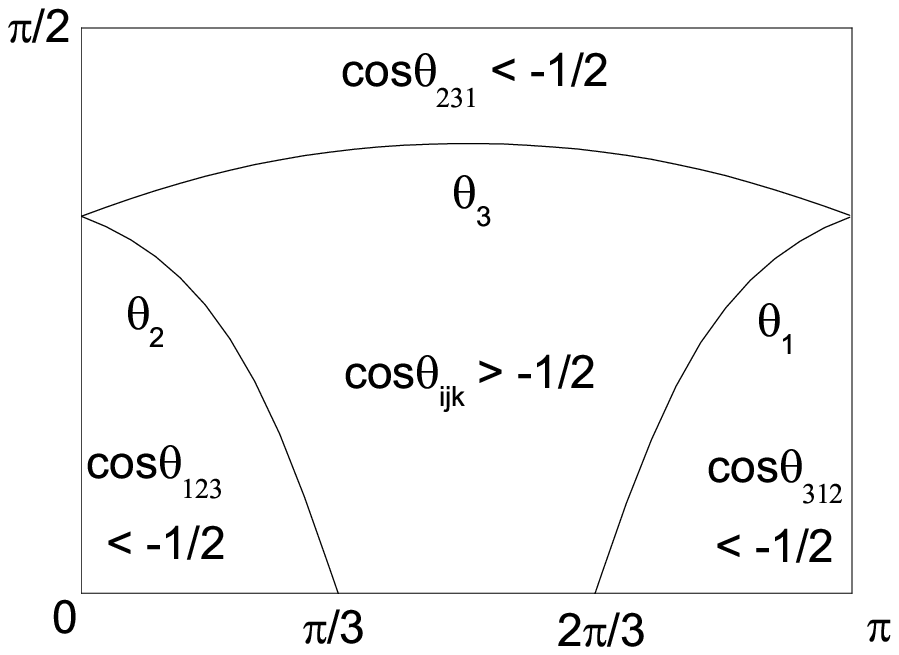}

\vspace*{15mm} {\Large\bfseries Fig.3.}
\end{center}

\newpage

\begin{center}
\includegraphics[width=160mm,
keepaspectratio=true]{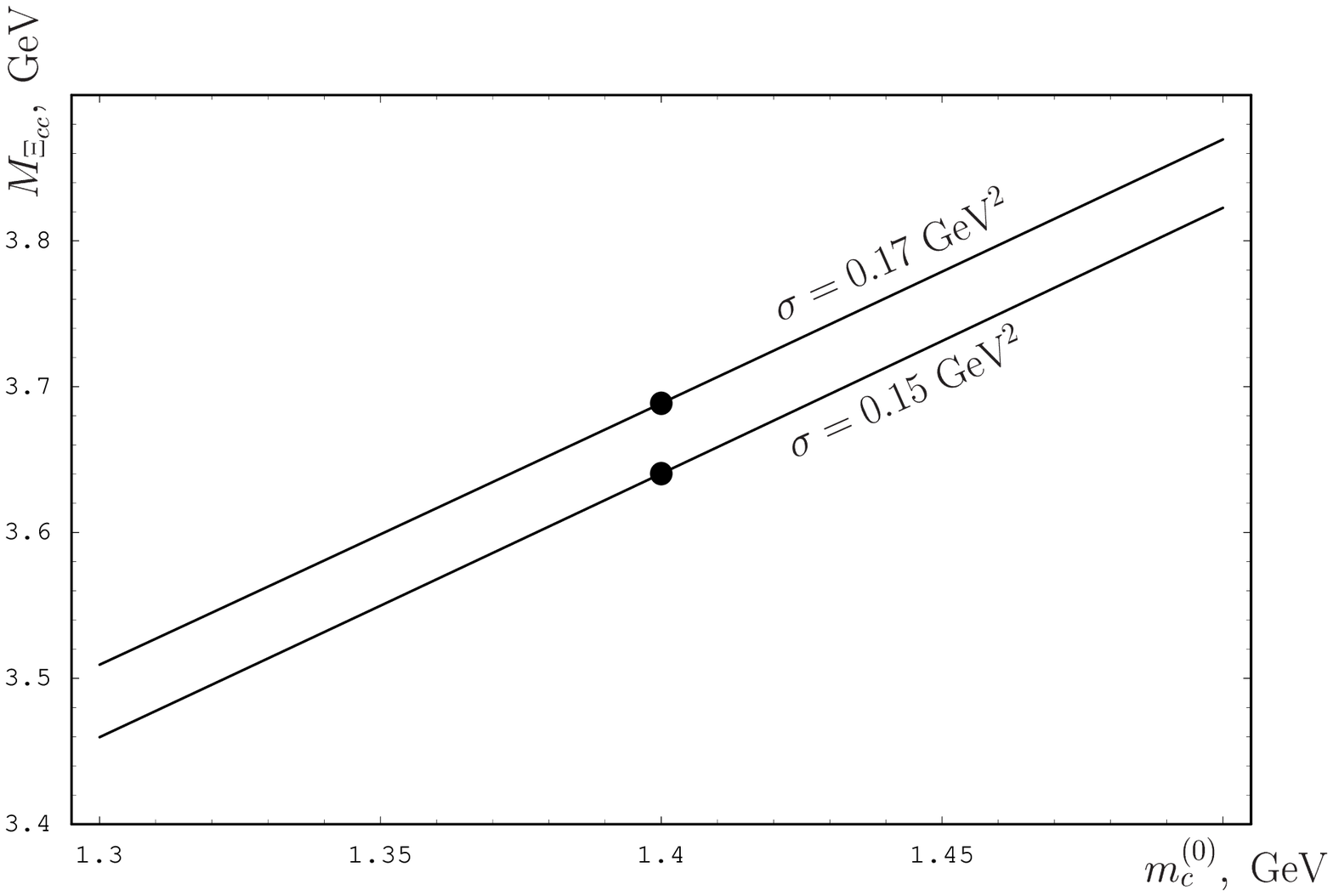}

\vspace*{15mm} {\Large\bfseries Fig.4.}
\end{center}


\newpage

\begin{thebibliography}{99}
\bibitem{ABE} F. Abe {\itshape et al.}, Phys. Rev. Lett. \textbf{81}, (1998) 2432,
hep-ex/9805034.

\bibitem{Mattson02} M. Mattson {\itshape et al.}, Phys. Rev. Lett. \textbf{89}, 112001
(2002).
\bibitem{KL02} V. V. Kiselev and A. K. Likhoded, hep-ph/0208231.
\bibitem{BaBar} see references [111--124] in  hep-ph/0201071.
\bibitem{NT01} I. M. Narodetskii and M. A. Trusov, Phys. Atom. Nucl. \textbf{65}, 917 (2002)
[hep-ph/0104019].
\bibitem{Si02}  Yu. A. Simonov, Phys. Atom. Nucl. \textbf{66}, 338 (2003) [hep-ph/0205334].
\bibitem{FS91} M. Fabre de la Ripelle and Yu. A. Simonov, Ann. Phys. (N.Y.) \textbf{212}, 235
(1991).
\bibitem{mesons1} A. Yu. Dubin, A. B. Kaidalov, and Yu. A. Simonov, Phys.
Lett. B \textbf{323}, 41 (1994).
\bibitem{mesons2} V. L. Morgunov, A. V. Nefediev, and Yu. A. Simonov, Phys.
Lett. B \textbf{459}, 653 (1999).
\bibitem{KS00} B. O. Kerbikov and Yu. A. Simonov, Phys. Rev. D \textbf{62},
093016 (2000).
\bibitem{Si01} Yu.A.Simonov, Phys. Lett. B \textbf{515}, 137
(2001).
\bibitem{ADM} X. Artru, Nucl. Phys. B \textbf{85}, 442 (1975); H. G. Dosch
and V. Mueller, Nucl. Phys. B \textbf{116}, 470 (1976).
\bibitem{CKP83} J. Carlson, J. Kogut, and V. R. Pandharipande,
Phys. Rev. D \textbf{27}, 233 (1983); N. Isgur and J. Paton, Phys.
Rev. D \textbf{31}, 2910 (1985).
\bibitem{CP02} S. Capstick and P. R. Page, Phys. Rev. C \textbf{66}, 065204
(2002) [hep-ph/0207027].
\bibitem{NT02} I. M. Narodetskii and M. A. Trusov, Nucl. Phys. B (Proc. Suppl.) \textbf{115}, 20
(2003) [hep-ph/0209044].
\bibitem{NPV02} I. M. Narodetskii, A. N. Plekhanov, and A. I. Veselov,
JETP Lett. \textbf{77}, 58 (2003) [hep-ph/0212235].
\bibitem{KNS87} Yu. S. Kalashnikova, I. M. Narodetskii, and Yu. A. Simonov, Yad. Fiz. \textbf{46}, 1181
(1987).
\bibitem{NT03} I. M. Narodetskii and M. A. Trusov, in \emph{Proceedings of
the 9th International Conference on the Structure of Baryons
(Baryons-2002), 3--8 March, 2002, Newport News, VA, USA}
[hep-ph/0304320].
\bibitem{TMNS02} T. T. Takahashi, H. Matsufuru, Y. Nemoto, and
H. Suganuma, Phys. Rev. Lett. \textbf{86}, 18 (2002).
\bibitem{KN00} Yu. S. Kalashnikova and A. Nefediev, Phys. Lett. B \textbf{492}, 91
(2000).
\bibitem{GKLO00} S. S. Gershtein, V. V. Kiselev, A. K. Likhoded,
and A. I. Onishchenko, Phys. Rev. D \textbf{62}, 050421 (2000).
\bibitem{EFGM02} D. Ebert, R. N. Faustov, V. D. Galkin, and
A. P. Matvienko, Phys. Rev. D \textbf{66}, 014008 (2002).
\bibitem{BDGNR94} E. Bagan {\itshape et al.}, Z. Phys. C \textbf{64}, 57
(1994).
\bibitem{deRuhula} A. De Rujula, H. Georgy, and S. L. Glashow, Phys. Rev. D \textbf{12}, 147
(1975).

\end{thebibliography}
\end{document}